\documentclass[leqno,final,notitlepage,a4paper,thmsa,onecolumn]{article}
\usepackage{amsfonts}
\usepackage{graphicx}

\usepackage{proc2e}
\usepackage[brazilian,american]{babel}

\input{tcilatex}

\begin{document}

\title{Quantum Battle of the Sexes Revisited}
\author{Jo\~{a}o Jos\'{e} de Farias Neto \\
%EndAName
Informatics Division\\
Institute for Advanced Studies\\
Centro T\'{e}cnico Aeroespacial, Brazil\\
joaojfn@ieav.cta.br}
\maketitle

\begin{abstract}
Several quantum versions of the battle of the sexes game are analyzed. Some
of them are shown to reproduce the classical game.\ In some, there are no
Nash quantum pure equilibria. In some others, the payoffs are always equal
to each other. In others still, all equilibria favor Alice or Bob depending
on a phase shift of the initial state of the system. Explicit detailed
calculations are for the first time exhibited.
\end{abstract}

\section{Introduction}

Since Meyer%
%TCIMACRO{\UNICODE{0xb4}}%
%BeginExpansion
\'{}%
%EndExpansion
s \cite{MEYER} idea of quantizing classical games, a flood of papers have
been published on this theme. Although that author%
%TCIMACRO{\UNICODE{0xb4}}%
%BeginExpansion
\'{}%
%EndExpansion
s motivation was to explore the mathematical resemblance between mixed
strategies and quantum superposition, the strongest argument in favor of
this line of research is the pending question about the possibility of
existence of quantum algorithms computationally more efficient then
classical ones for the solution of large sequential games (like chess, for
instance); research in this line is described in \cite{Cerf}, \cite{Farhi}, 
\cite{FarhiG}, \cite{Hogg} and \cite{Hogg98}. If such algorithms do exist,
they must be applied to a quantum version of the classical game, thus
requiring the fidelity of the former to the late one.

Several quantum versions for the battle of the sexes game have been proposed 
\cite{DUXULI}, \cite{DUXULI1}, \cite{Nawaz}, \cite{Marinatto}. Benjamin and
Hayden \cite{Benjamin} recommend Eisert%
%TCIMACRO{\UNICODE{0xb4}}%
%BeginExpansion
\'{}%
%EndExpansion
s \cite{EISWL} quantization scheme, which computes the final state as $%
|E_{f}\rangle =J^{\dagger }\left( U_{A}\otimes U_{B}\right) J|E_{i}\rangle $%
, because it generalizes correctly the classical game. Here, this last
framework is used. Firstly, the operator J is set equal to Eisert%
%TCIMACRO{\UNICODE{0xb4}}%
%BeginExpansion
\'{}%
%EndExpansion
s with $\gamma =\frac{\pi }{2}$ (which totally entangles the qubits) and
four cases are analyzed: (3,3), meaning unrestricted U$_{A}$ and U$_{B}$
(each one thus defined by three real numbers); (2,2), which is the
restricted version used by Eisert; (2,1), in which one of the players uses a
restricted quantum operator and the other can only use classical ones; and
(1,1), which corresponds to the classical game. Then, J is set to the
identity and the results for some initial states $|E_{i}\rangle $ are
obtained; the interesting \ cases here are when $|E_{i}\rangle $ is an
entangled state combining the two pure equilibria of the classical game.

\section{The classical game}

Alice and Bob are dating and want to go out together on Saturday night. But
Bob prefers to go to the football , whilst Alice prefers to go to the
ballet. Each one prefers to go with the other to the show (s)he doesn%
%TCIMACRO{\UNICODE{0xb4}}%
%BeginExpansion
\'{}%
%EndExpansion
t like instead of going alone to the show (s)he likes. The following table
may represent the personal satisfaction (payoff) of each player in the four
possible situations:

\ \ \ \ \ \ \ \ \ \ \ \ \ \ \ \ \ \ \ \ \ \ \ \ \ \ \ \ \ Bob

Alice$\left[ 
\begin{array}{ccc}
& Football(F) & Ballet(B) \\ 
Football & (1,2) & (0,0) \\ 
Ballet & (0,0) & (2,1)
\end{array}
\right] $

Lines represent Alice%
%TCIMACRO{\UNICODE{0xb4}}%
%BeginExpansion
\'{}%
%EndExpansion
s choices and columns, Bob%
%TCIMACRO{\UNICODE{0xb4}}%
%BeginExpansion
\'{}%
%EndExpansion
s. The first entry in each ordered pair is Alice%
%TCIMACRO{\UNICODE{0xb4}}%
%BeginExpansion
\'{}%
%EndExpansion
s payoff and the second one, Bob%
%TCIMACRO{\UNICODE{0xb4}}%
%BeginExpansion
\'{}%
%EndExpansion
s. The decisions have to be made simultaneously, independently and
thoroughly followed.

This game has two pure equilibria: FF and BB and a mixed one, in which Alice
chooses ballet with probability $\frac{2}{3}$ and Bob chooses football with
probability $\frac{2}{3}$.

\section{Quantization}

\bigskip

The two possible choices (F or B) can be associated with the direction (up
or down) of the spin of the nucleus of a carbon 13 atom, for instance,
embedded in a magnetic field. Suppose Alice and Bob control, each, one atom.
Two initial states are considered here (of the four possible ones): 
\TEXTsymbol{\vert}FF$\rangle $ and \TEXTsymbol{\vert}BB$\rangle $ ,
corresponding to the pure equilibria of the classical game. The judge
entangles the qubits with the aid of the operator

$J=e^{i\frac{\gamma }{2}C\otimes C}=e^{i\frac{\pi }{4}C\otimes C}=e^{i\frac{%
\pi }{4}\left[ 
\begin{array}{cc}
0 & 1 \\ 
-1 & 0
\end{array}
\right] \otimes \left[ 
\begin{array}{cc}
0 & 1 \\ 
-1 & 0
\end{array}
\right] }=$

$=e^{i\frac{\pi }{4}\left[ 
\begin{array}{cccc}
0 & 0 & 0 & 1 \\ 
0 & 0 & -1 & 0 \\ 
0 & -1 & 0 & 0 \\ 
1 & 0 & 0 & 0
\end{array}
\right] }=e^{\left[ 
\begin{array}{cccc}
0 & 0 & 0 & i\frac{\pi }{4} \\ 
0 & 0 & -i\frac{\pi }{4} & 0 \\ 
0 & -i\frac{\pi }{4} & 0 & 0 \\ 
i\frac{\pi }{4} & 0 & 0 & 0
\end{array}
\right] }$

$\allowbreak =\left[ 
\begin{array}{cccc}
\cos \frac{\pi }{4} & 0 & 0 & i\sin \frac{\pi }{4} \\ 
0 & \cos \frac{\pi }{4} & -i\sin \frac{\pi }{4} & 0 \\ 
0 & -i\sin \frac{\pi }{4} & \cos \frac{\pi }{4} & 0 \\ 
i\sin \frac{\pi }{4} & 0 & 0 & \cos \frac{\pi }{4}
\end{array}
\right] \allowbreak \allowbreak =$

$\allowbreak \left[ 
\begin{array}{cccc}
\frac{1}{2}\sqrt{2} & 0 & 0 & \frac{1}{2}i\sqrt{2} \\ 
0 & \frac{1}{2}\sqrt{2} & -\frac{1}{2}i\sqrt{2} & 0 \\ 
0 & -\frac{1}{2}i\sqrt{2} & \frac{1}{2}\sqrt{2} & 0 \\ 
\frac{1}{2}i\sqrt{2} & 0 & 0 & \frac{1}{2}\sqrt{2}
\end{array}
\right] \allowbreak $

Defining the basis of the system%
%TCIMACRO{\UNICODE{0xb4}}%
%BeginExpansion
\'{}%
%EndExpansion
s state space as $\left\{ |FF\rangle ,|FB\rangle ,|BF\rangle |BB\rangle
\right\} $, it follows that $|FF\rangle =(1,0,0,0),$

$|FB\rangle =(0,1,0,0),|BF\rangle =(0,0,1,0),|BB\rangle =(0,0,0,1)$. Thus, $%
J|FF\rangle =\left( \frac{1}{2}\sqrt{2},0,0,\frac{1}{2}i\sqrt{2}\right) $ and

$J|BB\rangle =\allowbreak \allowbreak \left( \frac{1}{2}i\sqrt{2},0,0,\frac{1%
}{2}\sqrt{2}\right) .$

Notice that $J|FF\rangle $ and $J|BB\rangle $ seem to be fair in the sense
that they assign equal probabilities ($\frac{1}{2}=\left| \frac{1}{2}i\sqrt{2%
}\right| ^{2}=\left| \frac{1}{2}\sqrt{2}\right| )$ to both pure equilibria
(FF and BB) of the classical game.

Now, Alice chooses a quantum operator U$_{A}$ and Bob, U$_{B}$, each one
acting only upon his own atom (qubit). As stated in the introduction, the
final state is computed by $|E_{f}\rangle =J^{\dagger }\left( U_{A}\otimes
U_{B}\right) J|E_{i}\rangle $ and the qubits are measured, thus collapsing
to one of the four basis states, corresponding to the four possible outcomes
of the classical game. If $|E_{f}\rangle =(r,s,t,u)$, meaning \ $%
|E_{f}\rangle =r|FF\rangle +s|FB\rangle +t|BF\rangle +u|BB\rangle $, the
expected payoffs are, then:

\$$_{A}=\left| r\right| ^{2}+2\left| u\right| ^{2}$

\$$_{B}=2\left| r\right| ^{2}+\left| u\right| ^{2}$

General quantum unitary operators acting over one qubit can be defined by

$U_{A}=\left[ 
\begin{array}{cc}
a & b \\ 
-b^{\ast } & a^{\ast }
\end{array}
\right] $

$U_{B}=\left[ 
\begin{array}{cc}
c & d \\ 
-d^{\ast } & c^{\ast }
\end{array}
\right] $

with $\left| a\right| ^{2}+\left| b\right| ^{2}=$ $\left| c\right|
^{2}+\left| d\right| ^{2}=1$ (see \cite{Marinatto}).

Now,

$U_{A}\otimes U_{B}=\left[ 
\begin{array}{cc}
a & b \\ 
-b^{\ast } & a^{\ast }
\end{array}
\right] \otimes \left[ 
\begin{array}{cc}
c & d \\ 
-d^{\ast } & c^{\ast }
\end{array}
\right] =\left[ 
\begin{array}{cc}
a\left[ 
\begin{array}{cc}
c & d \\ 
-d^{\ast } & c^{\ast }
\end{array}
\right] & b\left[ 
\begin{array}{cc}
c & d \\ 
-d^{\ast } & c^{\ast }
\end{array}
\right] \\ 
-b^{\ast }\left[ 
\begin{array}{cc}
c & d \\ 
-d^{\ast } & c^{\ast }
\end{array}
\right] & a^{\ast }\left[ 
\begin{array}{cc}
c & d \\ 
-d^{\ast } & c^{\ast }
\end{array}
\right]
\end{array}
\right] =\allowbreak \left[ 
\begin{array}{cccc}
ac & ad & bc & bd \\ 
-ad^{\ast } & ac^{\ast } & -bd^{\ast } & bc^{\ast } \\ 
-b^{\ast }c & -b^{\ast }d & a^{\ast }c & a^{\ast }d \\ 
b^{\ast }d^{\ast } & -b^{\ast }c^{\ast } & -a^{\ast }d^{\ast } & a^{\ast
}c^{\ast }
\end{array}
\right] $

So

$\left( U_{A}\otimes U_{B}\right) J|FF\rangle =\left[ 
\begin{array}{c}
\frac{1}{2}ac\sqrt{2}+\frac{1}{2}ibd\sqrt{2} \\ 
-\frac{1}{2}ad^{\ast }\sqrt{2}+\frac{1}{2}ibc^{\ast }\sqrt{2} \\ 
-\frac{1}{2}b^{\ast }c\sqrt{2}+\frac{1}{2}ia^{\ast }d\sqrt{2} \\ 
\frac{1}{2}b^{\ast }d^{\ast }\sqrt{2}+\frac{1}{2}ia^{\ast }c^{\ast }\sqrt{2}
\end{array}
\right] \allowbreak $

$\left( U_{A}\otimes U_{B}\right) J|BB\rangle =\allowbreak \allowbreak \left[
\begin{array}{c}
\frac{1}{2}iac\sqrt{2}+\frac{1}{2}bd\sqrt{2} \\ 
-\frac{1}{2}iad^{\ast }\sqrt{2}+\frac{1}{2}bc^{\ast }\sqrt{2} \\ 
-\frac{1}{2}ib^{\ast }c\sqrt{2}+\frac{1}{2}a^{\ast }d\sqrt{2} \\ 
\frac{1}{2}ib^{\ast }d^{\ast }\sqrt{2}+\frac{1}{2}a^{\ast }c^{\ast }\sqrt{2}
\end{array}
\right] \allowbreak \allowbreak $

\bigskip The Hermitian conjugate of J is

$J^{\dagger }=\left[ 
\begin{array}{cccc}
\frac{1}{2}\sqrt{2} & 0 & 0 & -\frac{1}{2}i\sqrt{2} \\ 
0 & \frac{1}{2}\sqrt{2} & \frac{1}{2}i\sqrt{2} & 0 \\ 
0 & \frac{1}{2}i\sqrt{2} & \frac{1}{2}\sqrt{2} & 0 \\ 
-\frac{1}{2}i\sqrt{2} & 0 & 0 & \frac{1}{2}\sqrt{2}
\end{array}
\right] $

As a consequence,

$J^{\dagger }\left( U_{A}\otimes U_{B}\right) J|FF\rangle =\left[ 
\begin{array}{c}
\frac{1}{2}ac+\frac{1}{2}ibd-\frac{1}{2}ib^{\ast }d^{\ast }+\frac{1}{2}%
a^{\ast }c^{\ast } \\ 
-\frac{1}{2}ad^{\ast }+\frac{1}{2}ibc^{\ast }-\frac{1}{2}ib^{\ast }c-\frac{1%
}{2}a^{\ast }d \\ 
-\frac{1}{2}iad^{\ast }-\frac{1}{2}bc^{\ast }-\frac{1}{2}b^{\ast }c+\frac{1}{%
2}ia^{\ast }d \\ 
-\frac{1}{2}iac+\frac{1}{2}bd+\frac{1}{2}b^{\ast }d^{\ast }+\frac{1}{2}%
ia^{\ast }c^{\ast }
\end{array}
\right] =\allowbreak \left[ 
\begin{array}{c}
\func{Re}(ac)-\func{Im}(bd) \\ 
\func{Im}(b^{\ast }c)-\func{Re}(a^{\ast }d) \\ 
\func{Im}(ad^{\ast })-\func{Re}(bc^{\ast }) \\ 
\func{Re}(bd)+\func{Im}(ac)
\end{array}
\right] $

And

$J^{\dagger }\left( U_{A}\otimes U_{B}\right) J|BB\rangle =\allowbreak \left[
\begin{array}{c}
\frac{1}{2}iac+\frac{1}{2}bd+\frac{1}{2}b^{\ast }d^{\ast }-\frac{1}{2}%
ia^{\ast }c^{\ast } \\ 
-\frac{1}{2}iad^{\ast }+\frac{1}{2}bc^{\ast }+\frac{1}{2}b^{\ast }c+\frac{1}{%
2}ia^{\ast }d \\ 
\frac{1}{2}ad^{\ast }+\frac{1}{2}ibc^{\ast }-\frac{1}{2}ib^{\ast }c+\frac{1}{%
2}a^{\ast }d \\ 
\frac{1}{2}ac-\frac{1}{2}ibd+\frac{1}{2}ib^{\ast }d^{\ast }+\frac{1}{2}%
a^{\ast }c^{\ast }
\end{array}
\right] \allowbreak =\left[ 
\begin{array}{c}
\func{Re}(bd)-\func{Im}(ac) \\ 
\func{Re}(bc^{\ast })+\func{Im}(ad^{\ast }) \\ 
\func{Re}(ad^{\ast })+\func{Im}(b^{\ast }c) \\ 
\func{Re}(ac)+\func{Im}(bd)
\end{array}
\right] $

$\allowbreak $

The most general form of $a,b,c,d$ is:

$a=e^{i\alpha }\cos \frac{1}{2}\theta $

$b=e^{i\beta }\sin \frac{1}{2}\theta $

$c=e^{i\gamma }\cos \frac{1}{2}\omega $

$d=e^{i\delta }\sin \frac{1}{2}\omega $

which gives the operators

$U_{A}=\left[ 
\begin{array}{cc}
e^{i\alpha }\cos \frac{1}{2}\theta & e^{i\beta }\sin \frac{1}{2}\theta \\ 
-e^{-i\beta }\sin \frac{1}{2}\theta & e^{-i\alpha }\cos \frac{1}{2}\theta
\end{array}
\right] \allowbreak $

$U_{B}=\allowbreak \left[ 
\begin{array}{cc}
e^{i\gamma }\cos \frac{1}{2}\omega & e^{i\delta }\sin \frac{1}{2}\omega \\ 
-e^{-i\delta }\sin \frac{1}{2}\omega & e^{-i\gamma }\cos \frac{1}{2}\omega
\end{array}
\right] $

Calling $|E_{FF}\rangle =J^{\dagger }\left( U_{A}\otimes U_{B}\right)
J|FF\rangle $ and $|E_{BB}\rangle =J^{\dagger }\left( U_{A}\otimes
U_{B}\right) J|BB\rangle $, the final states in function of these angles
become:

$|E_{FF}\rangle =\left[ 
\begin{array}{c}
\cos \frac{1}{2}\theta \cos \frac{1}{2}\omega \cos \left( \alpha +\gamma
\right) -\sin \frac{1}{2}\theta \sin \frac{1}{2}\omega \sin \left( \beta
+\delta \right) \\ 
\sin \frac{1}{2}\theta \cos \frac{1}{2}\omega \sin \left( -\beta +\gamma
\right) -\cos \frac{1}{2}\theta \sin \frac{1}{2}\omega \cos \left( \alpha
-\delta \right) \\ 
-\sin \frac{1}{2}\theta \cos \frac{1}{2}\omega \cos \left( -\beta +\gamma
\right) +\cos \frac{1}{2}\theta \sin \frac{1}{2}\omega \sin \left( \alpha
-\delta \right) \\ 
\cos \frac{1}{2}\theta \cos \frac{1}{2}\omega \sin \left( \alpha +\gamma
\right) +\sin \frac{1}{2}\theta \sin \frac{1}{2}\omega \cos \left( \beta
+\delta \right)
\end{array}
\right] \allowbreak $

$|E_{BB}\rangle =\left[ 
\begin{array}{c}
-\cos \frac{1}{2}\theta \cos \frac{1}{2}\omega \sin \left( \alpha +\gamma
\right) +\sin \frac{1}{2}\theta \sin \frac{1}{2}\omega \cos \left( \beta
+\delta \right) \\ 
\sin \frac{1}{2}\theta \cos \frac{1}{2}\omega \cos \left( -\beta +\gamma
\right) +\cos \frac{1}{2}\theta \sin \frac{1}{2}\omega \sin \left( \alpha
-\delta \right) \\ 
\sin \frac{1}{2}\theta \cos \frac{1}{2}\omega \sin \left( -\beta +\gamma
\right) +\cos \frac{1}{2}\theta \sin \frac{1}{2}\omega \cos \left( \alpha
-\delta \right) \\ 
\cos \frac{1}{2}\theta \cos \frac{1}{2}\omega \cos \left( \alpha +\gamma
\right) +\sin \frac{1}{2}\theta \sin \frac{1}{2}\omega \sin \left( \beta
+\delta \right)
\end{array}
\right] $

Defining

$x=\cos \frac{1}{2}\theta $

$y=\cos \frac{1}{2}\omega $

$z=\sin \left( \alpha +\gamma \right) $

$w=\sin \left( \beta +\delta \right) $

it is possible to write

$|E_{FF}\rangle =\left[ 
\begin{array}{c}
xy\sqrt{\left( 1-z^{2}\right) }-\sqrt{\left( 1-x^{2}\right) }\sqrt{\left(
1-y^{2}\right) }w \\ 
-\sqrt{\left( 1-x^{2}\right) }y\sin \left( \beta -\gamma \right) -x\sqrt{%
\left( 1-y^{2}\right) }\cos \left( -\alpha +\delta \right) \\ 
-\sqrt{\left( 1-x^{2}\right) }y\cos \left( \beta -\gamma \right) -x\sqrt{%
\left( 1-y^{2}\right) }\sin \left( -\alpha +\delta \right) \\ 
xyz+\sqrt{\left( 1-x^{2}\right) }\sqrt{\left( 1-y^{2}\right) }\sqrt{\left(
1-w^{2}\right) }
\end{array}
\right] \allowbreak $

$|E_{BB}\rangle =\left[ 
\begin{array}{c}
-xyz+\sqrt{\left( 1-x^{2}\right) }\sqrt{\left( 1-y^{2}\right) }\sqrt{\left(
1-w^{2}\right) } \\ 
\sqrt{\left( 1-x^{2}\right) }y\cos \left( \beta -\gamma \right) -x\sqrt{%
\left( 1-y^{2}\right) }\sin \left( -\alpha +\delta \right) \\ 
-\sqrt{\left( 1-x^{2}\right) }y\sin \left( \beta -\gamma \right) +x\sqrt{%
\left( 1-y^{2}\right) }\cos \left( -\alpha +\delta \right) \\ 
xy\sqrt{\left( 1-z^{2}\right) }+\sqrt{\left( 1-x^{2}\right) }\sqrt{\left(
1-y^{2}\right) }w
\end{array}
\right] $

The restriction $\beta =\delta =0$, used by some authors results on:

$|E_{FF}\rangle =\allowbreak \left[ 
\begin{array}{c}
xy\sqrt{\left( 1-z^{2}\right) } \\ 
-\sqrt{\left( 1-x^{2}\right) }y\sin \left( -\gamma \right) -x\sqrt{\left(
1-y^{2}\right) }\cos \left( -\alpha \right) \\ 
-\sqrt{\left( 1-x^{2}\right) }y\cos \left( -\gamma \right) -x\sqrt{\left(
1-y^{2}\right) }\sin \left( -\alpha \right) \\ 
xyz+\sqrt{\left( 1-x^{2}\right) }\sqrt{\left( 1-y^{2}\right) }
\end{array}
\right] \allowbreak $

$E_{BB}\rangle =\left[ 
\begin{array}{c}
-xyz+\sqrt{\left( 1-x^{2}\right) }\sqrt{\left( 1-y^{2}\right) } \\ 
\sqrt{\left( 1-x^{2}\right) }y\cos \left( -\gamma \right) -x\sqrt{\left(
1-y^{2}\right) }\sin \left( -\alpha \right) \\ 
-\sqrt{\left( 1-x^{2}\right) }y\sin \left( -\gamma \right) +x\sqrt{\left(
1-y^{2}\right) }\cos \left( -\alpha \right) \\ 
xy\sqrt{\left( 1-z^{2}\right) }
\end{array}
\right] $

Further restriction $\alpha =\gamma =0$ reduces to the classical case

$\allowbreak |E_{FF}\rangle =\left[ 
\begin{array}{c}
xy \\ 
-x\sqrt{\left( 1-y^{2}\right) } \\ 
-\sqrt{\left( 1-x^{2}\right) }y \\ 
\sqrt{\left( 1-x^{2}\right) }\sqrt{\left( 1-y^{2}\right) }
\end{array}
\right] \allowbreak $

$\allowbreak \allowbreak |E_{BB}\rangle =\allowbreak \left[ 
\begin{array}{c}
\sqrt{\left( 1-x^{2}\right) }\sqrt{\left( 1-y^{2}\right) } \\ 
\sqrt{\left( 1-x^{2}\right) }y- \\ 
x\sqrt{\left( 1-y^{2}\right) } \\ 
xy
\end{array}
\right] \allowbreak $

The asymmetric game, in which Alice can use restricted quantum operators and
Bob, only classical ones, is obtained by imposing $\gamma =0$ on the $%
2\times 2$ quantum case, resulting on:

$\bigskip |E_{FF}\rangle =\allowbreak \left[ 
\begin{array}{c}
\cos \frac{1}{2}\theta \cos \frac{1}{2}\omega \cos \alpha \\ 
-\cos \frac{1}{2}\theta \sin \frac{1}{2}\omega \cos \alpha \\ 
-\sin \frac{1}{2}\theta \cos \frac{1}{2}\omega +\cos \frac{1}{2}\theta \sin 
\frac{1}{2}\omega \sin \alpha \\ 
\cos \frac{1}{2}\theta \cos \frac{1}{2}\omega \sin \alpha +\sin \frac{1}{2}%
\theta \sin \frac{1}{2}\omega
\end{array}
\right] \allowbreak =\left[ 
\begin{array}{c}
xy\sqrt{\left( 1-z^{2}\right) } \\ 
-x\sqrt{\left( 1-y^{2}\right) }\sqrt{\left( 1-z^{2}\right) } \\ 
-\sqrt{\left( 1-x^{2}\right) }y+x\sqrt{\left( 1-y^{2}\right) }z \\ 
xyz+\sqrt{\left( 1-x^{2}\right) }\sqrt{\left( 1-y^{2}\right) }
\end{array}
\right] \allowbreak $

$|E_{BB}\rangle =\allowbreak \left[ 
\begin{array}{c}
-\cos \frac{1}{2}\theta \cos \frac{1}{2}\omega \sin \alpha +\sin \frac{1}{2}%
\theta \sin \frac{1}{2}\omega \\ 
\sin \frac{1}{2}\theta \cos \frac{1}{2}\omega +\cos \frac{1}{2}\theta \sin 
\frac{1}{2}\omega \sin \alpha \\ 
\cos \frac{1}{2}\theta \sin \frac{1}{2}\omega \cos \alpha \\ 
\cos \frac{1}{2}\theta \cos \frac{1}{2}\omega \cos \alpha
\end{array}
\right] \allowbreak =\allowbreak \left[ 
\begin{array}{c}
-xyz+\sqrt{\left( 1-x^{2}\right) }\sqrt{\left( 1-y^{2}\right) } \\ 
\sqrt{\left( 1-x^{2}\right) }y+x\sqrt{\left( 1-y^{2}\right) }z \\ 
x\sqrt{\left( 1-y^{2}\right) }\sqrt{\left( 1-z^{2}\right) } \\ 
xy\sqrt{\left( 1-z^{2}\right) }
\end{array}
\right] \allowbreak $

where $z=\sin \alpha $

\section{Analysis}

\subsection{$\allowbreak $Case 3$\times $3}

As remarked by Du et. al. \cite{DUXULI}, the totally entangled $3\times 3$
case doesn%
%TCIMACRO{\UNICODE{0xb4}}%
%BeginExpansion
\'{}%
%EndExpansion
t have pure quantum equilibria (mixed quantum equilibria, defined by way of
probability measures over the set of admissible density matrices, may exist
- in fact, Lee and Jonhson \cite{Lee} seem to have proved they always do).
The argument here goes as follows: define $u=xy=\cos \frac{1}{2}\theta \cos 
\frac{1}{2}\omega $ and $v=\sqrt{\left( 1-x^{2}\right) }\sqrt{\left(
1-y^{2}\right) }=\sin \frac{1}{2}\theta \sin \frac{1}{2}\omega $ ; then, the
payoffs corresponding to $|E_{BB}\rangle $ are

$\$_{A}=\left( -uz+v\sqrt{\left( 1-w^{2}\right) }\right) ^{2}+2\left( u\sqrt{%
\left( 1-z^{2}\right) }+vw\right) ^{2}\allowbreak $

$\$_{B}=2\left( -uz+v\sqrt{\left( 1-w^{2}\right) }\right) ^{2}+\left( u\sqrt{%
\left( 1-z^{2}\right) }+vw\right) ^{2}\allowbreak $

Let $z=\sin \rho $ and $w=\sin \sigma $. Then

$\$_{A}=\allowbreak \left( -u\sin \rho +v\sqrt{\left( 1-\sin ^{2}\sigma
\right) }\right) ^{2}+2\left( u\sqrt{\left( 1-\sin ^{2}\rho \right) }+v\sin
\sigma \right) ^{2}\allowbreak =\allowbreak \frac{3}{2}u^{2}+\frac{1}{2}%
u^{2}\cos 2\rho +uv\sin \left( \rho +\sigma \right) -3uv\sin \left( \rho
-\sigma \right) -\allowbreak \frac{1}{2}v^{2}\cos 2\sigma +\frac{3}{2}v^{2}$
, whose only candidates for extrema on the variables $\rho $ and $\sigma $
are $\allowbreak 2\left( u\pm v\right) ^{2}$ $,\allowbreak $ at $\left( \rho
,\sigma \right) \in \left\{ 0,k\frac{\pi }{2}|k\in \Bbb{Z}\right\} $. These
values for $\left( \rho ,\sigma \right) $ correspond to $\left( z,w\right)
\in \left\{ 0,\pm 1\right\} $. With such values of z and w, the results are
always $\$_{B}=2\$_{A}$ or $\$_{A}=2\$_{B}$, where $\min \left(
\$_{A},\$_{B}\right) \in \left\{ \left( u\pm v\right) ^{2}\right\} $. As
neither Alice nor Bob can control z or w independently, whatever be the
triplet ($\theta ,\alpha ,\beta )$ chosen by Alice, Bob can always change $%
\left( \gamma ,\delta \right) $ so as to set z and w to values that give him 
$\$_{B}=\max \left\{ \allowbreak 2\left( u+v\right) ^{2},\allowbreak 2\left(
u-v\right) ^{2}\right\} $ and, whatever be the triplet $\left( \omega
,\gamma ,\delta \right) $ chosen by Bob, Alice can always change $\left(
\alpha ,\beta \right) $ so as to set $z$ and $w$ to values that give her $%
\$_{A}=\max \left\{ \allowbreak 2\left( u+v\right) ^{2},\allowbreak 2\left(
u-v\right) ^{2}\right\} $. And both can escape the case $\$_{B}=\$_{A}=0$
(when $u=v=0$) by changing unilaterally the parameters that they really
control ($\theta $ for Alice, $\omega $ for Bob), as $u=v=0$ requires $%
\left( z=1,w=0\right) $ or $\left( z=0,w=1\right) .$

The same analysis applies to $|E_{FF}\rangle $.

\subsection{Case 2$\times $2}

The payoffs in this case are:

$\$_{A}(FF)=\left( xy\sqrt{\left( 1-z^{2}\right) }\right) ^{2}+2\left( xyz+%
\sqrt{\left( 1-x^{2}\right) }\sqrt{\left( 1-y^{2}\right) }\right) ^{2}$

$\$_{B}(FF)=2\left( xy\sqrt{\left( 1-z^{2}\right) }\right) ^{2}+\left( xyz+%
\sqrt{\left( 1-x^{2}\right) }\sqrt{\left( 1-y^{2}\right) }\right) ^{2}$

$\$_{A}(BB)=\left( -xyz+\sqrt{\left( 1-x^{2}\right) }\sqrt{\left(
1-y^{2}\right) }\right) ^{2}+2\left( xy\sqrt{\left( 1-z^{2}\right) }\right)
^{2}$

$\$_{B}(BB)=2\left( -xyz+\sqrt{\left( 1-x^{2}\right) }\sqrt{\left(
1-y^{2}\right) }\right) ^{2}+\left( xy\sqrt{\left( 1-z^{2}\right) }\right)
^{2}$

\bigskip

where $(FF)$ and $(BB)$ indicate the initial state $|E_{i}\rangle $ of the
system. It is sufficient to notice that, in the case FF, Alice can always
annulate the first term by setting $x=0$, which results in $\$_{A}=2\$_{B}$,
giving Bob no other choice then agreeing to maximize the second term, which
is accomplished by setting $y=0$, resulting in $\$_{B}=1$. As Alice can
never expect to get a payoff bigger then 2, the resulting profile gives no
incentive to change for the members of the couple, thus being a Nash
equilibrium. In the case BB, it is Bob who has the \textit{couteau} and the 
\textit{fromage,} being able to annulate the second term, by setting $y=0$,
leaving Alice no choice other then choosing $x=0$ to maximize her payoff.

Reminding that $x=\cos \frac{1}{2}\theta $ \ and $y=\cos \frac{1}{2}\omega ,$%
in all of the above equilibria the players use the flipping operators:

$U_{A}=\left[ 
\begin{array}{cc}
e^{i\alpha }\cos \frac{1}{2}\theta & \sin \frac{1}{2}\theta \\ 
-\sin \frac{1}{2}\theta & e^{-i\alpha }\cos \frac{1}{2}\theta
\end{array}
\right] \allowbreak =\allowbreak \left[ 
\begin{array}{cc}
0 & \pm 1 \\ 
\mp 1 & 0
\end{array}
\right] $

$U_{B}=\allowbreak \left[ 
\begin{array}{cc}
e^{i\gamma }\cos \frac{1}{2}\omega & \sin \frac{1}{2}\omega \\ 
-\sin \frac{1}{2}\omega & e^{-i\gamma }\cos \frac{1}{2}\omega
\end{array}
\right] =\allowbreak \left[ 
\begin{array}{cc}
0 & \pm 1 \\ 
\mp 1 & 0
\end{array}
\right] $

Indeed all equilibria of this case favor Alice, when $|E_{i}\rangle
=|FF\rangle $ and Bob, when $|E_{i}\rangle =|BB\rangle .$

To see this, write

$\$_{A}(FF)=u^{2}\left( 1-z^{2}\right) +2\left( uz+v\right) ^{2}=\allowbreak
u^{2}z^{2}+4vuz+\left( u^{2}+2v^{2}\right) $

$\$_{B}(FF)=2u^{2}\left( 1-z^{2}\right) +\left( uz+v\right) ^{2}=\allowbreak
-u^{2}z^{2}+2vuz+2u^{2}+v^{2}$

\bigskip with the same definitions of $u$ and $v$ as before. Then

$\frac{\partial }{\partial z}\$_{A}(FF)=\allowbreak 2u^{2}z+4vu$

$\frac{\partial }{\partial z}\$_{B}(FF)=\allowbreak -2u^{2}z+2vu$

$\frac{\partial ^{2}}{\partial z^{2}}\$_{A}(FF)=\allowbreak 2u^{2}$

$\frac{\partial ^{2}}{\partial z^{2}}\$_{B}(FF)=\allowbreak -2u^{2}$

Alice%
%TCIMACRO{\UNICODE{0xb4}}%
%BeginExpansion
\'{}%
%EndExpansion
s payoff is maximized at the extremities $\left| z\right| =1$; Bob%
%TCIMACRO{\UNICODE{0xb4}}%
%BeginExpansion
\'{}%
%EndExpansion
s, on $z=\frac{v}{u}$. As neither can control z independently, whatever
value one sets to z the other one can unset. As a consequence, in any
equilibrium, they must agree as to the best value of z. This will be the
case only if $\left| \frac{v}{u}\right| =1$. Then, the payoffs are reduced
to $\$_{A}(FF)=2\$_{B}(FF)=2\left( uz+v\right) ^{2}$, obliging them to
coordinate to make $\left| uz+v\right| =1$, thus jointly maximizing their
payoffs, but maintaining the bias towards Alice, who gets 2, while Bob gets
1. Besides, Bob must not have incentive to choose $y$ such that $\left| 
\frac{v}{u}\right| \neq 1$.

Equilibria of the kind $\left| z\right| =1,\left| u\right| =\left| v\right| $
reduce the payoffs to

$\$_{A}=2\left( uz+v\right) ^{2}\,$

$\$_{B}=\left( uz+v\right) ^{2}$

If $z=1$, then $\left| u+v\right| =1$ and $v=u=\pm \frac{1}{2}$.

If $z=-1$, then $\left| -u+v\right| =1$ and $v=-u=\pm \frac{1}{2}$.

Now,

$u=xy=\frac{1}{2}$ implies $y=\frac{1}{2x}$

$v=\sqrt{1-x^{2}}\sqrt{1-y^{2}}=\allowbreak \frac{1}{2}\sqrt{\left(
1-x^{2}\right) }\sqrt{\left( 4-\frac{1}{x^{2}}\right) }=\allowbreak \frac{1}{%
2}\frac{\sqrt{\left( 1-x^{2}\right) }}{x}\sqrt{\left( 4x^{2}-1\right) }$

\bigskip $\allowbreak \frac{1}{2}\frac{\sqrt{\left( 1-x^{2}\right) }}{x}%
\sqrt{\left( 4x^{2}-1\right) }=\allowbreak \frac{1}{2}$, which results on $x=%
\frac{1}{2}\sqrt{2}$ and, thus, $y=\frac{1}{2}\sqrt{2}$

As a consequence, these equilibria require $\left| z\right| =1$, $\left|
x\right| =\frac{1}{2}\sqrt{2}=\left| y\right| =\frac{1}{2}\sqrt{2}$.

The condition $\left| x\right| =\frac{1}{2}\sqrt{2}$ is translated for $%
\theta $ as:

$\left| \cos \frac{1}{2}\theta \right| =\frac{1}{2}\sqrt{2}$, which is
satisfied by $\theta =\pm \allowbreak \left( \frac{1}{2}\pi +2k\pi \right) $

By the same token, $\omega =\pm \allowbreak \left( \frac{1}{2}\pi +2k\pi
\right) $

The corresponding operators are of the type:

$U_{A}=\frac{1}{2}\sqrt{2}\left[ 
\begin{array}{cc}
\pm e^{i\alpha } & \pm 1 \\ 
\mp 1 & \pm e^{-i\alpha }
\end{array}
\right] \allowbreak $

$U_{B}=\frac{1}{2}\sqrt{2}\left[ 
\begin{array}{cc}
\pm e^{i\gamma } & \pm 1 \\ 
\mp 1 & \pm e^{-i\gamma }
\end{array}
\right] \allowbreak $

recalling that these equilibria, as the former ones, give 2 to Alice and 1
to Bob.

Given the symmetry, the analysis of $|E_{BB}\rangle $ follows the same
lines. In this case, all equilibria give 2 to Bob and 1 to Alice.

What is new here, besides the explicitation of the analysis, is the
comparison between the two cases, FF and BB, showing that the simple
interchange of $i=e^{i\frac{\pi }{2}}$ in the components of the state $%
J|E_{i}\rangle $, changes completely the balance of a game otherwise
symmetrical. That is: define the system being controlled by the players by
the pair ($J,|G_{i}\rangle )$, where $|G_{i}\rangle =J|E_{i}\rangle $; then,
a change from $|G_{i}\rangle =\left( \frac{1}{2}\sqrt{2},0,0,\frac{1}{2}i%
\sqrt{2}\right) $ to $|G_{i}\rangle =\left( \frac{1}{2}i\sqrt{2},0,0,\frac{1%
}{2}\sqrt{2}\right) $ turns a game totally favorable to Alice to one totally
favorable to Bob.

\subsection{\protect\bigskip Case 2$\times $1}

\bigskip

In this case, the operators available for the players are:

$U_{A}=\left[ 
\begin{array}{cc}
e^{i\alpha }\cos \frac{\theta }{2} & \sin \frac{\theta }{2} \\ 
-\sin \frac{\theta }{2} & e^{-i\alpha }\cos \frac{\theta }{2}
\end{array}
\right] $

$U_{B}=\allowbreak \left[ 
\begin{array}{cc}
\cos \frac{1}{2}\omega & \sin \frac{1}{2}\omega \\ 
-\sin \frac{1}{2}\omega & \cos \frac{1}{2}\omega
\end{array}
\right] $

Then,

$\allowbreak $

$\$_{A}(FF)=\left( \cos \frac{1}{2}\theta \cos \frac{1}{2}\omega \cos \alpha
\right) ^{2}+2\left( \cos \frac{1}{2}\theta \cos \frac{1}{2}\omega \sin
\alpha +\sin \frac{1}{2}\theta \sin \frac{1}{2}\omega \right) ^{2}$

$\$_{B}(FF)=2\left( \cos \frac{1}{2}\theta \cos \frac{1}{2}\omega \cos
\alpha \right) ^{2}+\left( \cos \frac{1}{2}\theta \cos \frac{1}{2}\omega
\sin \alpha +\sin \frac{1}{2}\theta \sin \frac{1}{2}\omega \right) ^{2}$

$\$_{A}(BB)=\left( -\cos \frac{1}{2}\theta \cos \frac{1}{2}\omega \sin
\alpha +\sin \frac{1}{2}\theta \sin \frac{1}{2}\omega \right) ^{2}+2\left(
\cos \frac{1}{2}\theta \cos \frac{1}{2}\omega \cos \alpha \right) ^{2}$

$\$_{B}(BB)=2\left( -\cos \frac{1}{2}\theta \cos \frac{1}{2}\omega \sin
\alpha +\sin \frac{1}{2}\theta \sin \frac{1}{2}\omega \right) ^{2}+\left(
\cos \frac{1}{2}\theta \cos \frac{1}{2}\omega \cos \alpha \right) ^{2}$

With the same definitions as before, the only difference being that now $%
z=\sin \alpha $, which gives it total control by Alice, the analysis is the
following:

The first and fourth components \ of the final quantum state $|E_{FF}\rangle 
$ are:

$\left[ 
\begin{array}{c}
xy\sqrt{\left( 1-z^{2}\right) } \\ 
xyz+\sqrt{\left( 1-x^{2}\right) }\sqrt{\left( 1-y^{2}\right) }
\end{array}
\right] $

The payoffs become, with the definitions of u and v already given:

$\$_{A}=u^{2}(1-z^{2})+2(uz+v)^{2}=\allowbreak u^{2}z^{2}+4vuz+u^{2}+2v^{2}$

$\$_{B}=2u^{2}(1-z^{2})+(uz+v)^{2}=\allowbreak -u^{2}z^{2}+2vuz+2u^{2}+v^{2}$

As $\$_{A}(z)$ is a parable upwards, Alice always chooses $\left| z\right|
=1.$ Then,

$\$_{A}=2(uz+v)^{2}$

$\$_{B}=(uz+v)^{2}$

Alice e Bob coordinate to maximize $\left| uz+v\right| .$

Because the minimum of \$$_{A}(z)$ is at $z=-\frac{2v}{u}$, if u and v have
opposite signs, Alice chooses z=-1; otherwise, z=1. Now,

$u=\cos \frac{1}{2}\theta \cos \frac{1}{2}\omega $

$v=\sin \frac{1}{2}\theta \sin \frac{1}{2}\omega $

Suppose c and d have equal signs. Then, z=1 and

$\left| uz+v\right| =\left| u+v\right| =\allowbreak \left| \cos \frac{1}{2}%
\theta \cos \frac{1}{2}\omega +\sin \frac{1}{2}\theta \sin \frac{1}{2}\omega
\right| $

The maximum value of this expression is 1, attained a $\omega =\theta +2k\pi
,$ $k\in \Bbb{Z}$.

Suppose c and d have opposite signs. Then, z=-1 and

$\left| uz-v\right| =\left| u-v\right| =\allowbreak \left| \cos \frac{1}{2}%
\theta \cos \frac{1}{2}\omega -\sin \frac{1}{2}\theta \sin \frac{1}{2}\omega
\right| $

The maximum value of this expression is 1, attained a $\omega =-\theta
+2k\pi ,$ $k\in \Bbb{Z}$.

Particularly, for $\theta ,\omega \in \left\{ \pi +2k\pi |k\in \Bbb{Z}%
\right\} $, U$_{A}$ and U$_{B}$ become independent from $\alpha $, so that
this parameter is free.

When $|E_{i}\rangle =|BB\rangle $, the final state has the following first
and fourth components:

$\left[ 
\begin{array}{c}
-xyz+\sqrt{\left( 1-x^{2}\right) }\sqrt{\left( 1-y^{2}\right) } \\ 
xy\sqrt{\left( 1-z^{2}\right) }
\end{array}
\right] $

The payoffs are:

$\$_{A}=2u^{2}(1-z^{2})+(-uz+v)^{2}=\allowbreak
-u^{2}z^{2}-2vuz+2u^{2}+v^{2} $

$\$_{B}=u^{2}(1-z^{2})+2(uz+v)^{2}=\allowbreak u^{2}z^{2}+4vuz+u^{2}+2v^{2}$

$\frac{\partial }{\partial z}\$_{A}=\allowbreak -2u^{2}z-2uv$

$\$_{A}(z)$ is a downwards parable, with its maximum at $z=-\frac{1}{u}v.$

So, if $\frac{1}{u}v\in \left[ -1,1\right] $, Alice chooses $z=-\frac{1}{u}v$%
; otherwise, $\left| z\right| =1.$

Suppose $\frac{1}{u}v\in \left[ -1,1\right] $. Then,

$\$_{A}=2u^{2}\left( 1-\frac{1}{u^{2}}v^{2}\right) +4v^{2}=\allowbreak
2u^{2}+2v^{2}$

$\$_{B}=u^{2}\left( 1-\frac{1}{u^{2}}v^{2}\right) =\allowbreak u^{2}-v^{2}$

To maximize his payoff, Bob annulates $v$, by setting $\left| y\right| =1$.\
The resulting payoffs are:

$\$_{A}=\allowbreak 2u^{2}$

$\$_{B}=\allowbreak u^{2}$

and Alice maximizes u by choosing $\left| x\right| =1.$ As $v=0$ and $u\neq
0 $, the hypothesis $\frac{1}{u}v\in \left[ -1,1\right] $ is still valid.

This leaves Bob with no other choice than setting $|y|=1$. The result is 2
to Alice and 1 to Bob.

Suppose that $\frac{1}{u}v\notin \left[ -1,1\right] $. Then Alice chooses $%
\left| z\right| =1$ and

$\$_{A}=(-uz+v)^{2}$

$\$_{B}=2(uz+v)^{2}$

Thus, Bob tries to maximize $\left| uz+v\right| $ and Alice, $\left|
-uz+v\right| .$ Both are simultaneously maximized only when $uz=0$. As z$%
\neq 0$, u must be annulated, which any player has the power to do
unilaterally. So, in the equilibria, u=0 and the payoffs become

$\$_{A}=v^{2}$

$\$_{B}=\allowbreak 2v^{2}$

The players coordinate to maximize $\left| v\right| $ by setting $\left|
x\right| =\left| y\right| =1$. As $v\neq 0$ and $u=0$, the hypothesis $\frac{%
1}{u}v\notin \left[ -1,1\right] $ is still valid.

As u=0, z becomes irrelevant, which frees $\alpha .$ The result is 2 to Bob
and 1 to Alice.

This case, having only three real parameters, allows the graphical
illustration below, showing the sets of Nash equilibria in the parameter
space. \FRAME{ftbpF}{192.75pt}{135.1875pt}{0pt}{}{}{Figure }{\special%
{language "Scientific Word";type "GRAPHIC";maintain-aspect-ratio
TRUE;display "USEDEF";valid_file "T";width 192.75pt;height 135.1875pt;depth
0pt;original-width 189.6875pt;original-height 132.5pt;cropleft "0";croptop
"1";cropright "1";cropbottom "0";tempfilename
'FIG1.jpg';tempfile-properties "XPR";}}

Figure 1: Quantum Alice versus classical Bob - points of the parameter space
belonging to the bold straight segments are the Nash equilibria of this
quantum version of the battle of the sexes game. The initial state is FF.
All equilibria favor Alice (\$$_{A}=2$, \$$_{B}=1$). Vertical axis is $%
\omega $, horizontal is $\theta $ and the one perpendicular to the page is $%
\alpha $.

\FRAME{ftbpF}{201.8125pt}{136pt}{0pt}{}{}{Figure }{\special{language
"Scientific Word";type "GRAPHIC";maintain-aspect-ratio TRUE;display
"USEDEF";valid_file "T";width 201.8125pt;height 136pt;depth
0pt;original-width 198.75pt;original-height 133.25pt;cropleft "0";croptop
"1";cropright "1";cropbottom "0";tempfilename
'FIG2.jpg';tempfile-properties "XPR";}}

Figure 2: Quantum Alice versus classical Bob - points of the parameter space
belonging to the bold straight segments are Nash equilibria of this quantum
version of the battle of the sexes game. The initial state is BB. Most
equilibria favor Bob (\$$_{A}=1$, \$$_{B}=2$). In the isolate points shown,
the equilibrium favors Alice (\$$_{A}=2$, \$$_{B}=1$). Vertical axis is $%
\omega $, horizontal is $\theta $ and the one parallel to the bold straight
lines is $\alpha $.

\subsection{Some simpler models}

Marinatto and Weber \cite{Marinatto} analyzed the non-entangled case, where $%
J=I$ (the identity operator, obtained by setting $\gamma =0$ in J%
%TCIMACRO{\UNICODE{0xb4}}%
%BeginExpansion
\'{}%
%EndExpansion
s formula) with initial state $|BB\rangle $, without restrictions on the
parameters set. The final state is given by

$|E_{BB}\rangle =\left( U_{A}\otimes U_{B}\right) |BB\rangle =\left[ 
\begin{array}{cccc}
ac & ad & bc & bd \\ 
-ad^{\ast } & ac^{\ast } & -bd^{\ast } & bc^{\ast } \\ 
-b^{\ast }c & -b^{\ast }d & a^{\ast }c & a^{\ast }d \\ 
b^{\ast }d^{\ast } & -b^{\ast }c^{\ast } & -a^{\ast }d^{\ast } & a^{\ast
}c^{\ast }
\end{array}
\right] \left[ 
\begin{array}{c}
0 \\ 
0 \\ 
0 \\ 
1
\end{array}
\right] =\allowbreak \left[ 
\begin{array}{c}
bd \\ 
bc^{\ast } \\ 
a^{\ast }d \\ 
a^{\ast }c^{\ast }
\end{array}
\right] \allowbreak $

which results in the payoffs:

$\$_{A}=\left| bd\right| ^{2}+2|a^{\ast }c^{\ast }|^{2}=\left| b\right|
^{2}|d|^{2}+2\left| a\right| ^{2}\left| c\right| ^{2}=\left( 1-x^{2}\right)
\left( 1-y^{2}\right) +2x^{2}y^{2}$

$\$_{B}=2\left| bd\right| ^{2}+|a^{\ast }c^{\ast }|^{2}=2\left| b\right|
^{2}|d|^{2}+\left| a\right| ^{2}\left| c\right| ^{2}=2\left( 1-x^{2}\right)
\left( 1-y^{2}\right) +x^{2}y^{2}$

If the initial state is $|FF\rangle $, the situation changes to:

$|E_{FF}\rangle =\left( U_{A}\otimes U_{B}\right) |FF\rangle =\left[ 
\begin{array}{cccc}
ac & ad & bc & bd \\ 
-ad^{\ast } & ac^{\ast } & -bd^{\ast } & bc^{\ast } \\ 
-b^{\ast }c & -b^{\ast }d & a^{\ast }c & a^{\ast }d \\ 
b^{\ast }d^{\ast } & -b^{\ast }c^{\ast } & -a^{\ast }d^{\ast } & a^{\ast
}c^{\ast }
\end{array}
\right] \left[ 
\begin{array}{c}
1 \\ 
0 \\ 
0 \\ 
0
\end{array}
\right] =\allowbreak \left[ 
\begin{array}{c}
ac \\ 
-ad^{\ast } \\ 
-b^{\ast }c \\ 
b^{\ast }d^{\ast }
\end{array}
\right] \allowbreak $

which results in the payoffs:

$\$_{A}=2\left| b^{\ast }d^{\ast }\right| ^{2}+|ac|^{2}=2\left| b\right|
^{2}|d|^{2}+\left| a\right| ^{2}\left| c\right| ^{2}=2\left( 1-x^{2}\right)
\left( 1-y^{2}\right) +x^{2}y^{2}$

$\$_{B}=\left| b^{\ast }d^{\ast }\right| ^{2}+2|ac|^{2}=\left| b\right|
^{2}|d|^{2}+2\left| a\right| ^{2}\left| c\right| ^{2}=\left( 1-x^{2}\right)
\left( 1-y^{2}\right) +2x^{2}y^{2}$

\bigskip

Both cases above reproduce the classical game, sufficing to identify $x$ and 
$y$ with the probabilities of Alice and Bob, respectively, choosing ballet,
in the first case ($|E_{BB}\rangle $) and choosing football, in the second
one ($|E_{FF}\rangle $). Thus, the equilibria are $\left( x,y\right) \in
\left\{ \left( 1,1\right) ,\left( 0,0\right) ,\left( \frac{2}{3},\frac{1}{3}%
\right) \right\} $, in the first case, and $\left( x,y\right) \in \left\{
\left( 1,1\right) ,\left( 0,0\right) ,\left( \frac{1}{3},\frac{2}{3}\right)
\right\} $.

Now, suppose the initial state fed to the system is already entangled.

First, let $|E_{i}\rangle =\frac{\sqrt{2}}{2}\left( |FF\rangle +|BB\rangle
\right) .$

In the basis here used, this is represented by the vector $\frac{\sqrt{2}}{2}%
\left( 1001\right) $. So

$|E_{f}\rangle =\left( U_{A}\otimes U_{B}\right) |E_{i}\rangle =\frac{\sqrt{2%
}}{2}\left[ 
\begin{array}{cccc}
ac & ad & bc & bd \\ 
-ad^{\ast } & ac^{\ast } & -bd^{\ast } & bc^{\ast } \\ 
-b^{\ast }c & -b^{\ast }d & a^{\ast }c & a^{\ast }d \\ 
b^{\ast }d^{\ast } & -b^{\ast }c^{\ast } & -a^{\ast }d^{\ast } & a^{\ast
}c^{\ast }
\end{array}
\right] \left[ 
\begin{array}{c}
1 \\ 
0 \\ 
0 \\ 
1
\end{array}
\right] =\left[ 
\begin{array}{c}
\frac{1}{2}\sqrt{2}ac+\frac{1}{2}\sqrt{2}bd \\ 
-\frac{1}{2}\sqrt{2}ad^{\ast }+\frac{1}{2}\sqrt{2}bc^{\ast } \\ 
-\frac{1}{2}\sqrt{2}b^{\ast }c+\frac{1}{2}\sqrt{2}a^{\ast }d \\ 
\frac{1}{2}\sqrt{2}b^{\ast }d^{\ast }+\frac{1}{2}\sqrt{2}a^{\ast }c^{\ast }
\end{array}
\right] \allowbreak $

The resulting payoffs are:

$\$_{A}=\frac{1}{2}\left( \left| ac+bd\right| ^{2}+2|a^{\ast }c^{\ast
}+b^{\ast }d^{\ast }|^{2}\right) =\frac{3}{2}\left| ac+bd\right| ^{2}$

$\$_{B}=\frac{1}{2}\left( 2\left| ac+bd\right| ^{2}+|a^{\ast }c^{\ast
}+b^{\ast }d^{\ast }|^{2}\right) =\frac{3}{2}\left| ac+bd\right| ^{2}$

So, in this case, $\$_{A}=\$_{B}$ whatever are the values of the parameters
and this is a game of coordination, with both players being interested in
raising $\left| ac+bd\right| ^{2}$ to it%
%TCIMACRO{\UNICODE{0xb4}}%
%BeginExpansion
\'{}%
%EndExpansion
s maximal attainable value, which is 1. When they get this, the expected
payoffs are $\frac{3}{2}$ for each one. This is equivalent to, in the
classical game, the couple making a decision like a single entity, by
flipping a coin and deciding to go to football, if it turns out as head, and
to ballet otherwise, which is basically a one player (the couple) game.

Is there an asymmetry between $|E_{i}\rangle =\frac{\sqrt{2}}{2}\left(
i|FF\rangle +|BB\rangle \right) $ and $|E_{i}\rangle =\frac{\sqrt{2}}{2}%
\left( |FF\rangle +i|BB\rangle \right) $, as in the case when J was chosen
with $\gamma =\frac{\pi }{2}$?

This case is equivalent to defining the quantum game by $|E_{f}\rangle
=\left( U_{A}\otimes U_{B}\right) J|BB\rangle $ and $|E_{f}\rangle =\left(
U_{A}\otimes U_{B}\right) J|FF\rangle $, respectively, with $J=J\left(
\gamma =\frac{\pi }{2}\right) $, that is, skipping the post multiplication
by $J^{\dagger }$.

Suppose $|E_{i}\rangle =\frac{\sqrt{2}}{2}\left( i|FF\rangle +|BB\rangle
\right) $. Then

$|E_{f}\rangle =\left( U_{A}\otimes U_{B}\right) |E_{i}\rangle =\frac{\sqrt{2%
}}{2}\left[ 
\begin{array}{cccc}
ac & ad & bc & bd \\ 
-ad^{\ast } & ac^{\ast } & -bd^{\ast } & bc^{\ast } \\ 
-b^{\ast }c & -b^{\ast }d & a^{\ast }c & a^{\ast }d \\ 
b^{\ast }d^{\ast } & -b^{\ast }c^{\ast } & -a^{\ast }d^{\ast } & a^{\ast
}c^{\ast }
\end{array}
\right] \left[ 
\begin{array}{c}
i \\ 
0 \\ 
0 \\ 
1
\end{array}
\right] =\left[ 
\begin{array}{c}
\frac{1}{2}i\sqrt{2}ac+\frac{1}{2}\sqrt{2}bd \\ 
-\frac{1}{2}i\sqrt{2}ad^{\ast }+\frac{1}{2}\sqrt{2}bc^{\ast } \\ 
-\frac{1}{2}i\sqrt{2}b^{\ast }c+\frac{1}{2}\sqrt{2}a^{\ast }d \\ 
\frac{1}{2}i\sqrt{2}b^{\ast }d^{\ast }+\frac{1}{2}\sqrt{2}a^{\ast }c^{\ast }
\end{array}
\right] \allowbreak $

The resulting payoffs are:

$\$_{A}=\frac{1}{2}\left( \left| iac+bd\right| ^{2}+2|a^{\ast }c^{\ast
}+ib^{\ast }d^{\ast }|^{2}\right) $

$\$_{B}=\frac{1}{2}\left( 2\left| iac+bd\right| ^{2}+|a^{\ast }c^{\ast
}+ib^{\ast }d^{\ast }|^{2}\right) $

But $iac+bd=i\left( a^{\ast }c^{\ast }+ib^{\ast }d^{\ast }\right) ^{\ast }$

So $\left| iac+bd\right| ^{2}=|a^{\ast }c^{\ast }+ib^{\ast }d^{\ast }|^{2}$

As a result, $\$_{A}=\$_{B}=\frac{3}{2}\left| iac+bd\right| ^{2}$

and, again, the game is one of coordination, with all equilibria at $\left|
iac+bd\right| =1$ and $\$_{A}=\$_{B}=\frac{3}{2}$.

Suppose $|E_{i}\rangle =\frac{\sqrt{2}}{2}\left( |FF\rangle +i|BB\rangle
\right) $. Then

$|E_{f}\rangle =\left( U_{A}\otimes U_{B}\right) |E_{i}\rangle =\frac{\sqrt{2%
}}{2}\left[ 
\begin{array}{cccc}
ac & ad & bc & bd \\ 
-ad^{\ast } & ac^{\ast } & -bd^{\ast } & bc^{\ast } \\ 
-b^{\ast }c & -b^{\ast }d & a^{\ast }c & a^{\ast }d \\ 
b^{\ast }d^{\ast } & -b^{\ast }c^{\ast } & -a^{\ast }d^{\ast } & a^{\ast
}c^{\ast }
\end{array}
\right] \left[ 
\begin{array}{c}
1 \\ 
0 \\ 
0 \\ 
i
\end{array}
\right] =\left[ 
\begin{array}{c}
\frac{1}{2}\sqrt{2}ac+\frac{1}{2}\sqrt{2}ibd \\ 
-\frac{1}{2}\sqrt{2}ad^{\ast }+\frac{1}{2}\sqrt{2}ibc^{\ast } \\ 
-\frac{1}{2}\sqrt{2}b^{\ast }c+\frac{1}{2}\sqrt{2}ia^{\ast }d \\ 
\frac{1}{2}\sqrt{2}b^{\ast }d^{\ast }+\frac{1}{2}\sqrt{2}ia^{\ast }c^{\ast }
\end{array}
\right] \allowbreak $

The resulting payoffs are:

$\$_{A}=\frac{1}{2}\left( \left| ac+ibd\right| ^{2}+2|ia^{\ast }c^{\ast
}+b^{\ast }d^{\ast }|^{2}\right) $

$\$_{B}=\frac{1}{2}\left( 2\left| ac+ibd\right| ^{2}+|ia^{\ast }c^{\ast
}+b^{\ast }d^{\ast }|^{2}\right) $

But $ac+ibd=i\left( ia^{\ast }c^{\ast }+b^{\ast }d^{\ast }\right) ^{\ast }$

So $\left| ac+bd\right| ^{2}=|ia^{\ast }c^{\ast }+b^{\ast }d^{\ast }|^{2}$

As a result, $\$_{A}=\$_{B}=\frac{3}{2}\left| ac+ibd\right| ^{2}\allowbreak $

and, again, the game is one of coordination, with all equilibria at $\left|
ac+ibd\right| =1$ and $\$_{A}=\$_{B}=\frac{3}{2}$.

Paraphrasing Benjamin and Hayden \cite{Benjamin} , \textbf{in these coherent
equilibria, entanglement shared among the players enables different kinds of
cooperative behavior: indeed it can act as a contract, in the sense that it
obliges the players to coordinate with each other. }

\section{Conclusion}

Some versions of the quantum battle of the sexes game reproduce the
classical one, in the sense that equal proportions of the equilibria favor
Alice ($\$_{A}=2,\$_{B}=1$) and Bob ($\$_{A}=1,\$_{B}=2$) and another set
gives them both an expected payoff of $\frac{2}{3}$. In other versions, all
equilibria favor Alice or all favor Bob, the difference between one type and
the other being just a phase in the initial state of the system. In a third
kind of versions, the payoffs are equal to each other independently of the
operator that each player chooses, all equilibria giving both an expected
payoff of $\frac{3}{2}$. In a fourth kind, there are no Nash pure quantum
equilibria at all.

Quantum games, as any other game, can be viewed as a dispute between two or
more players to control a system; thus, they are defined by the system and
the way the players are connected to it; if any of these change, the game
changes; as a consequence, it is arguable if its name should be kept. To
name ''battle of the sexes'' games with completely different qualitative
features may not be reasonable.

\end{document}